\documentclass[12pt,preprint]{aastex}

\usepackage{emulateapj5}

\usepackage{graphicx}
\usepackage{amsmath}
\usepackage{times}

\shorttitle{QPOs in the Light Curves from Oscillating Tori}
\shortauthors{Schnittman \& Rezzolla}

\begin{document}

\title{Quasi-periodic Oscillations in the X-ray Light Curves from 
Relativistic Tori}

\author{Jeremy D.\ Schnittman\altaffilmark{1} and Luciano
Rezzolla\altaffilmark{2,~}\altaffilmark{3}}

\altaffiltext{1}{Department of Physics, Massachusetts Institute of Technology,
77 Massachusetts Avenue, Cambridge, MA 02139, USA}

\altaffiltext{2}{SISSA, International School for Advanced Studies and
INFN, Via Beirut, 2 34014 Trieste, Italy}

\altaffiltext{3}{Department of Physics and Astronomy, Louisiana State
University, 202 Nicholson Hall, Baton Rouge, LA 70803, USA} 

\begin{abstract}
	We  use a  relativistic  ray-tracing code  to  analyze the  X-ray
	emission from a pressure-supported oscillating relativistic torus
	around a  black hole.  We  show that a strong  correlation exists
	between the {\it intrinsic} frequencies of the torus normal modes
	and the  {\it extrinsic} frequencies  seen in the  observed light
	curve   power  spectrum.    This  correlation   demonstrates  the
	feasibility  of  the   oscillating-torus  model  to  explain  the
	multiple peaks  seen in black  hole high-frequency quasi-periodic
	oscillations.   Using an  optically thin,  monochromatic emission
	model,  we  also   determine  how  a  relativistically  broadened
	emission  line and  the  amplitude of  the  X-ray modulations are
	dependent on  the observer's inclination  angle and on  the torus
	oscillation  amplitudes.  Observations   of  these  features  can
	provide important information about the torus as well as the black
	hole.
\end{abstract}


\maketitle

\section{INTRODUCTION}
\label{intro}

	Recent observations with the {\it Rossi X-ray Timing Explorer}
({\it RXTE}) have revealed the existence of high-frequency quasi-periodic
oscillations (QPOs) in a number of accreting black hole binary
systems~\citep{stroh01,mccli05}. In an increasing number of these
systems, the QPOs appear with integer commensurabilities, generally a
$2:3$ frequency ratio~\citep{mille01,remil02,homan05}. Since these
modulations are expected to originate very close to the black hole, they
could be used to test gravity in strong-field regimes or extract
information on the black hole properties. 

	Over the years, a large number
of theoretical models have been developed to explain these observations.
Some of the more popular models explain the QPOs through magnetic
flares~\citep{galee79,pouta99}, fluid oscillations in thin
disks~\citep{okaza87,nowak97}, geodesic resonances
\citep{stell99,abram01}, and trapped fluid oscillations in tori around
black holes~\citep{rezzo03a,rezzo03b,montero04,lee04,zanot05}.

	To evaluate the relative strengths and weaknesses of any of these
models, it is essential to compare directly the predictions of the
theoretical models with the observations. To this end, we have applied
the ray-tracing methods described in~\citet{schni04} to a numerical
calculation of the nonlinear dynamics of a relativistic axisymmetric
torus described in~\citet{zanot03}. With this combined approach we have
calculated the profile of a relativistically broadened emission line,
the X-ray light curves, and the power spectra, all as measured by a distant 
observer. We have demonstrated that a strong correlation exists between
the intrinsic normal-mode oscillations of a pressure-supported torus and
the extrinsic observables of the X-ray light curves and power spectra. It
should be noted that this approach is rather different from the one recently
presented by~\citet{bursa04}, where the torus was modeled analytically
and thus not the result of self-consistent relativistic-hydrodynamics
simulations.

This Letter is organized as follows: in Section~\ref{radtranseqn} we
summarize the basic dynamical features of the perturbed relativistic tori
and describe how we apply the classical radiative transfer equation to a
general relativistic accretion model. In Section~\ref{results} we present
simulated images of the torus along with instantaneous line profiles and
integrated X-ray light curves for a range of inclinations and
perturbation amplitudes, representing the torus initial conditions. We
conclude in Section~\ref{discussion} with a discussion of the major
results and a look towards future work.

\section{
Torus Dynamics and Radiative Transfer}
\label{radtranseqn}

	To briefly summarize the basic properties of the
oscillating-torus model, we recall that we are considering a
non-self-gravitating perfect-fluid torus orbiting a Schwarzschild
black hole~\citep{font02,zanot03}. The fluid is assumed to be in circular
non-geodesic motion and the conditions of hydrostatic equilibrium and of
azimuthal symmetry allow the relativistic hydrodynamics equations to be reduced
to Bernoulli-type equations. These have particularly simple solutions
when the fluid is assumed to have a constant specific angular momentum
and if a polytropic equation of state is adopted. In this case, the
equations of hydrostatic equilibrium can be integrated analytically to
yield the rest-mass density distribution inside the torus, with the
isobaric surfaces coinciding with the equipotential ones. Hereafter we
will consider tori with constant specific angular momentum and bear in
mind that more complex distributions introduce only small quantitative
differences [see \citet{montero04, zanot05} for details].

	Once a stationary equilibrium configuration is constructed, it is
perturbed with the introduction of a small radial velocity expressed in
terms of the radial velocity for a relativistic spherically symmetric
accretion flow onto a Schwarzschild black hole, i.e.\ the Michel
solution~\citep{miche72}.  More specifically, we set the initial radial
(covariant) component of the fluid 3-velocity as $v_r = \eta (v_{r})_{_{\rm
Michel}}$, and then use the dimensionless coefficient $\eta$ to tune the
strength of the perturbation, obtaining an essentially linear response
for $\eta\lesssim 0.06$~\citep{zanot03}. As noted in~\citet{zanot05}, the
response of the torus is largely independent of the type of perturbation
and different choices lead to the excitation of the same modes.

	With these initial conditions, the equations of relativistic
hydrodynamics in a Schwarzschild black hole space-time are solved using
the axisymmetric, general relativistic code described
in~\citet{zanot03} and~\citet{zanot05}. This makes use of a first-order,
flux-conservative formulation of the equations, which are solved using a
high-resolution shock-capturing scheme based on an approximate Riemann
solver. Second-order accuracy in both space and time is achieved by
adopting a piecewise-linear cell reconstruction procedure and a
second-order, conservative Runge-Kutta scheme, respectively. As the
numerical evolution proceeds, the density and pressure in
the fluid's local rest-frame, as well as the coordinate 4-velocity, are
tabulated and stored at each point in space-time.

	We recall that the introduction of the perturbations triggers
harmonic oscillations of the torus having centrifugal and
pressure-gradients as the restoring forces. A careful investigation of
these oscillations has also revealed that there are multiple peaks in the
power spectrum with frequencies in a sequence of integers:
2\;:\;3\;:\;4\;:...~\citep{zanot03,zanot05}. Subsequent perturbative
analyses have also shown that these oscillations are indeed $p$ modes,
behaving as trapped waves within the cavity produced by the torus and
hence having eigenfrequencies in a sequence of small
integers~\citep{rezzo03b,montero04}. The striking analogy between the
harmonic relation among the $p$-mode eigenfrequencies and the QPOs
observed in black-hole systems has then led to the suggestion that QPOs
could result from basic fluid oscillations of a small accretion torus
close to the black hole~\citep{rezzo03a}. The existence of such tori
appears to be a robust feature of global magnetohydrodynamic
simulations \citep{devil03}.

	While attractive for its simplicity and for being based on
global modes of oscillations that are expected to be present in
realistic accretion discs, the relativistic torus model has so far
only suggested a property of an orbiting, pressure-supported gas
itself. It is not intuitively obvious that ``intrinsic'' modulations
in the fluid hydrodynamics will produce similar ``extrinsic'' modulations
in the observed light curve. Furthermore, it is not clear {\it a
priori} what might be the relationship between the phases and
amplitudes of the intrinsic and extrinsic oscillations, and how these
might compare with the X-ray data.

	To explore this relationship and demonstrate how an oscillating
accretion model produces a corresponding oscillating signature in the
observed X-ray light curve, we have calculated the trajectories of
photons from a distant observer to the emission region around a black
hole following the methods described in~\citet{schni04}. In this
approach, the photon positions and momenta are tabulated
along each ray's path in order to recreate a simulated,
time-varying image of the accreting gas. Given the photon's 4-position
and 4-momentum along the entire path, the observed
spectrum is calculated for that ray by integrating the radiative transfer
equation with a model for the gas emissivity and absorption.

	More precisely, we start from the classical radiative transfer
equation~\citep{rybic79}
\begin{equation}\label{rad_trans_eq}
\frac{dI_\nu}{ds} = j_\nu -\alpha_\nu I_\nu\ ,
\end{equation}
where $ds$ is the differential path length and $I_\nu$, $j_\nu$, and
$\alpha_\nu$ are respectively the spectral intensity, emissivity, and
absorption coefficient of the fluid at a frequency $\nu$.
These variables are typically
defined in the rest-frame of the gas as functions of its local
temperature and density. We have explored a number of different emission
models, which will be presented in greater detail in a companion paper.
For simplicity, the results discussed here are primarily based on an
optically thin gas emitting isotropically and monochromatically at
frequency $\nu_{\rm em}$ with $\alpha_\nu=0$ and $j_\nu \propto
\rho\delta(\nu-\nu_{\rm em})$.

	We then incorporate relativistic effects by defining a local
orthonormal tetrad at each point along the integration path (for the
Schwarzschild geometry, this is simply the coordinate stationary observer
frame). Once transformed to this locally-flat tetrad, only {\it special}
relativistic effects must be included. Following~\citet{rybic79},
equation (\ref{rad_trans_eq}) becomes
\begin{equation}
\label{rad_trans_eq2}
\frac{dI_\nu}{ds} = \left(\frac{\nu}{\nu'}\right)^2 j_\nu' -
	\left(\frac{\nu'}{\nu}\right) \alpha_\nu' I_\nu \ ,
\end{equation}
where primed and unprimed variables are measured in the rest-frame of
the gas and in the stationary tetrad, respectively. The special
relativistic Doppler shift between the photon path and the fluid
can be written as
\begin{equation}
\frac{\nu'}{\nu} = \gamma(1-\beta\cos\theta)\ ,
\end{equation}
where $\beta \equiv v/c$, $\gamma \equiv 1/\sqrt{1-\beta^2}$, and
$\theta$ is the angle between the photon direction and the gas, all
measured in the stationary tetrad. In addition, as the photon bundle
propagates through the global curvature around the black hole, the
spectral intensity at a given frequency evolves as the photons are
gravitationally red-shifted, maintaining the Lorentz invariance of
$I_\nu/\nu^3$.

\section{LIGHT CURVES AND POWER SPECTRA}
\label{results}
\vbox{ 
\vskip 0.125truecm
\centerline{\epsfxsize=8truecm 
\epsfbox{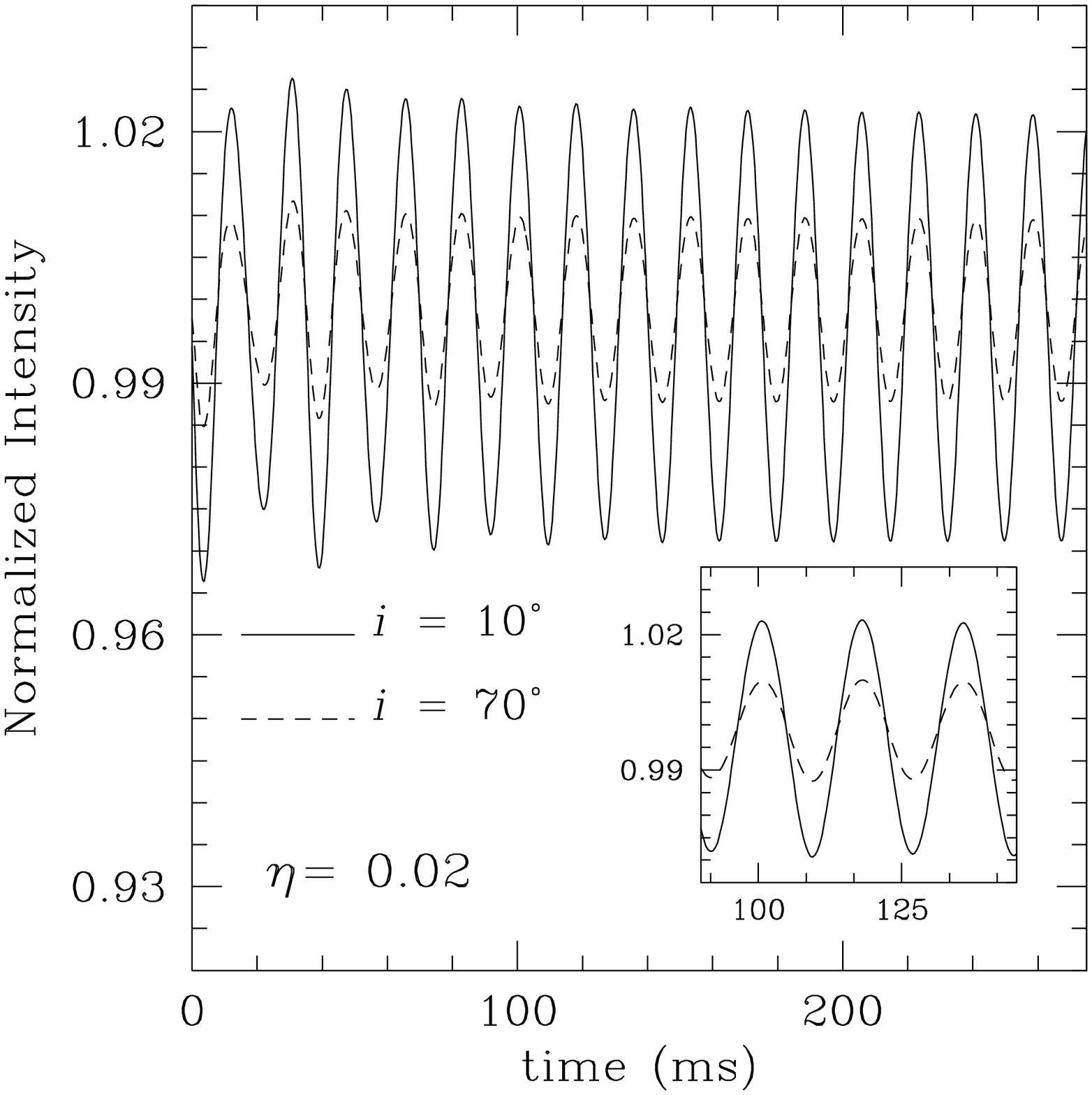}}
\figcaption[]{\label{plotone}Normalized X-ray light curves from
oscillating tori with inclinations $i=10^{\circ}$ ({\it solid}) and
$70^{\circ}$ ({\it dashed}), for a perturbation amplitude of
$\eta=0.02$. The local minima in the light curves correspond to minima in
the torus size, when it is closer to the black hole. The inset
shows the same light curves in greater detail.}
\vskip 0.125truecm
}

\begin{figure*}
\hskip 0.9truecm
\scalebox{0.4}{\includegraphics*[0,80][360,300]{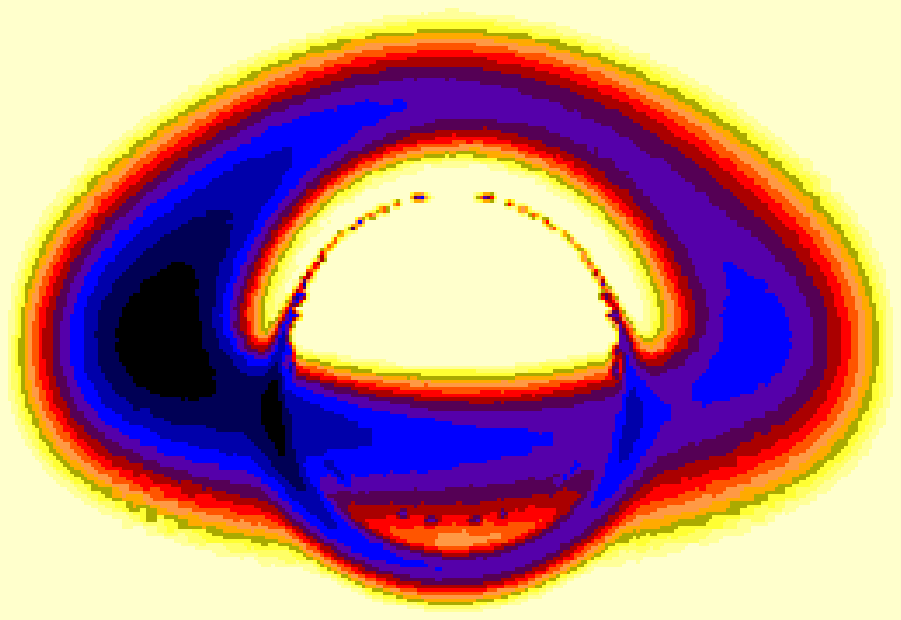}}
\hskip 0.9truecm
\scalebox{0.4}{\includegraphics*[0,80][360,300]{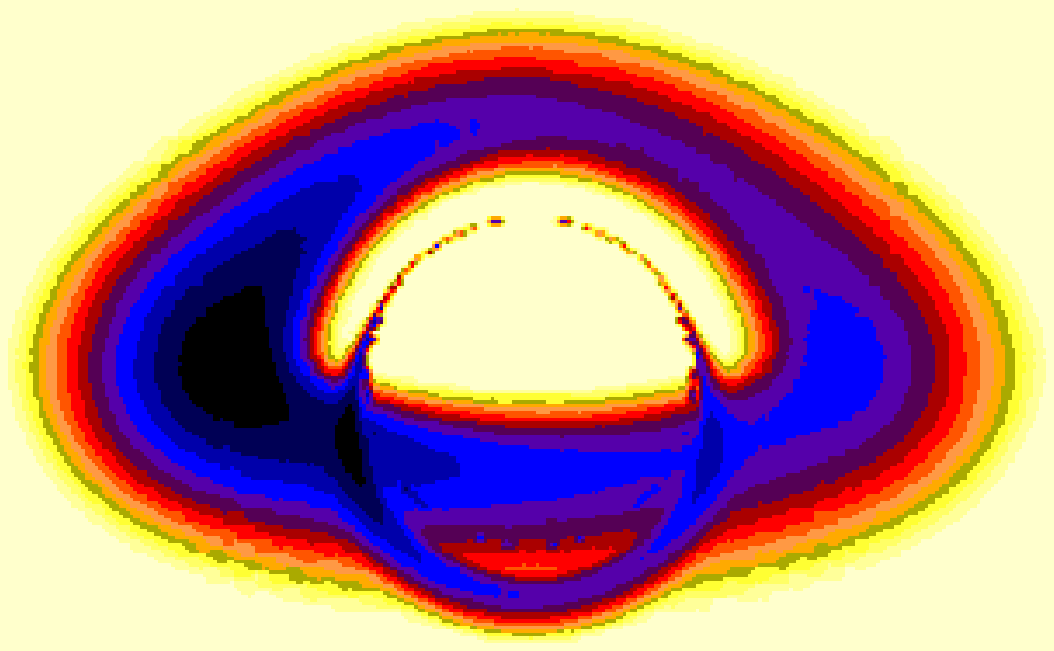}}
\hskip 0.9truecm
\scalebox{0.4}{\includegraphics*[0,80][360,300]{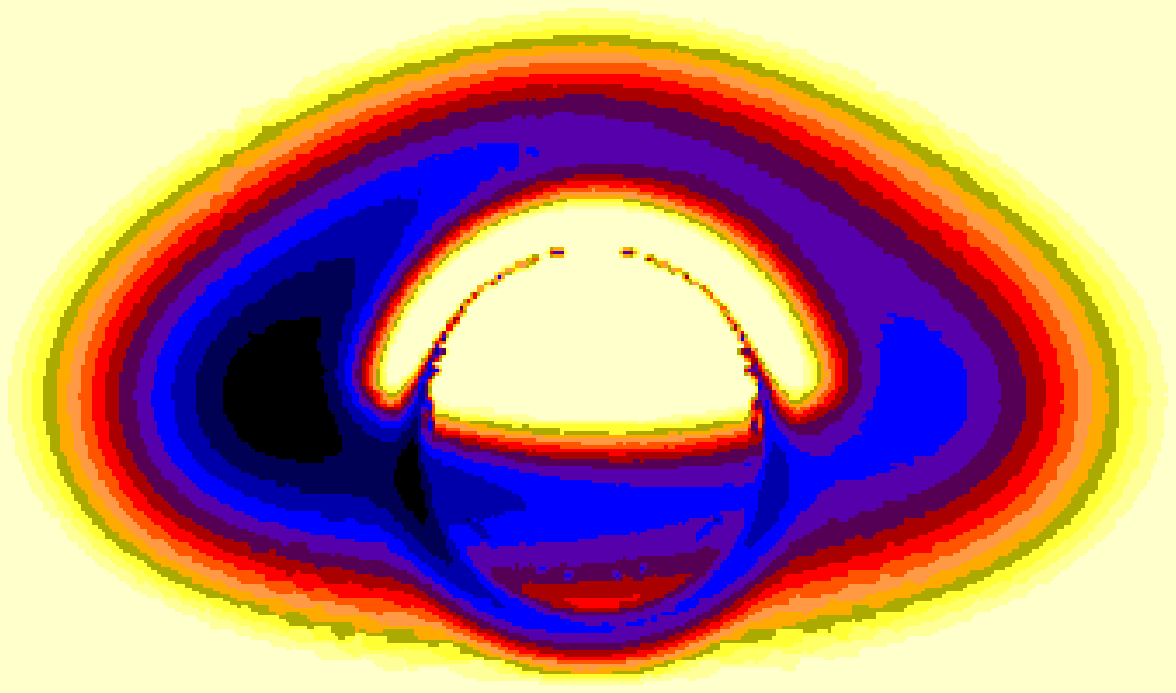}}
\vskip -1.2truecm
\hskip 0.9truecm
\includegraphics*[width=5.cm]{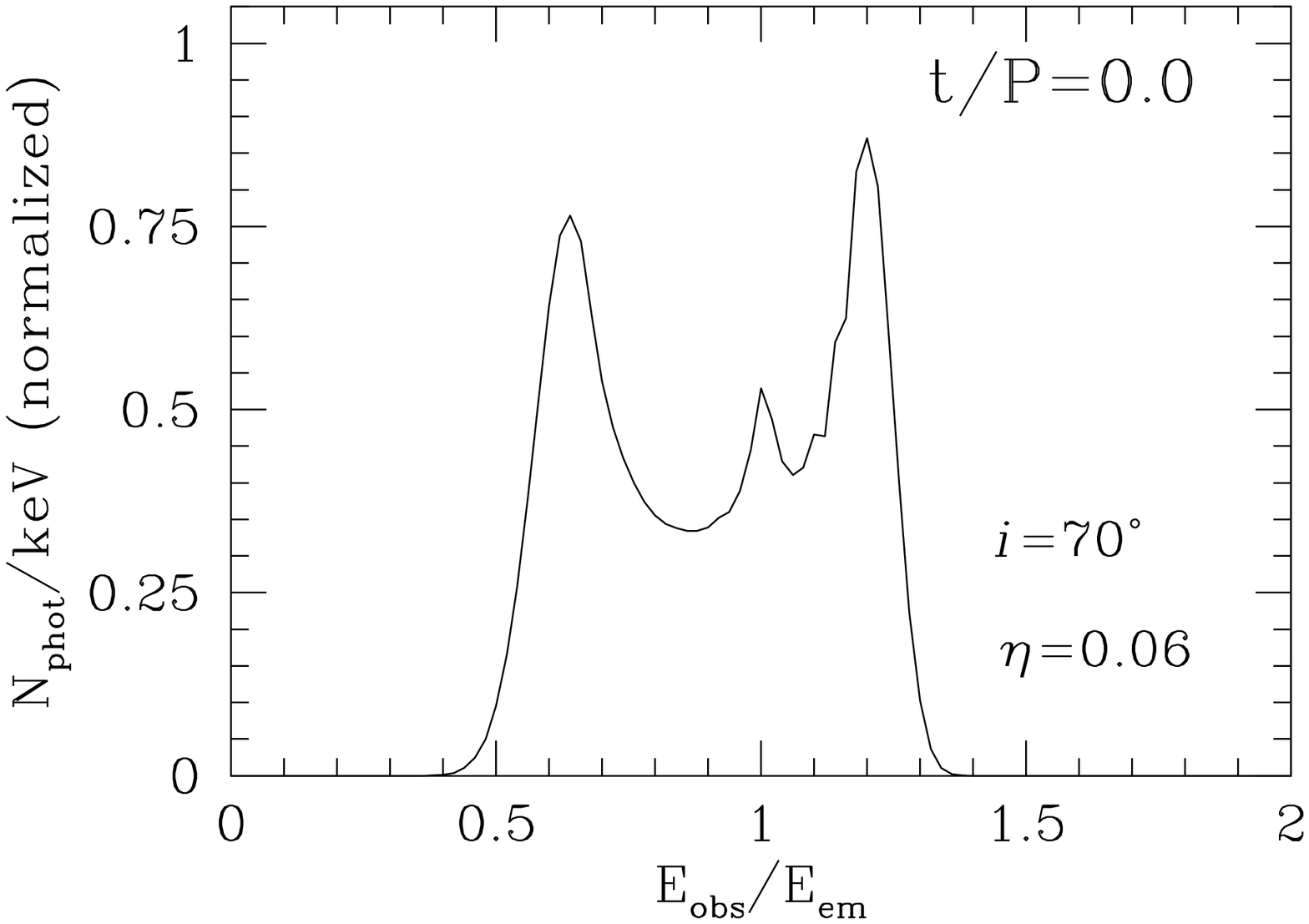}
\hskip 0.9truecm
\includegraphics*[width=5.cm]{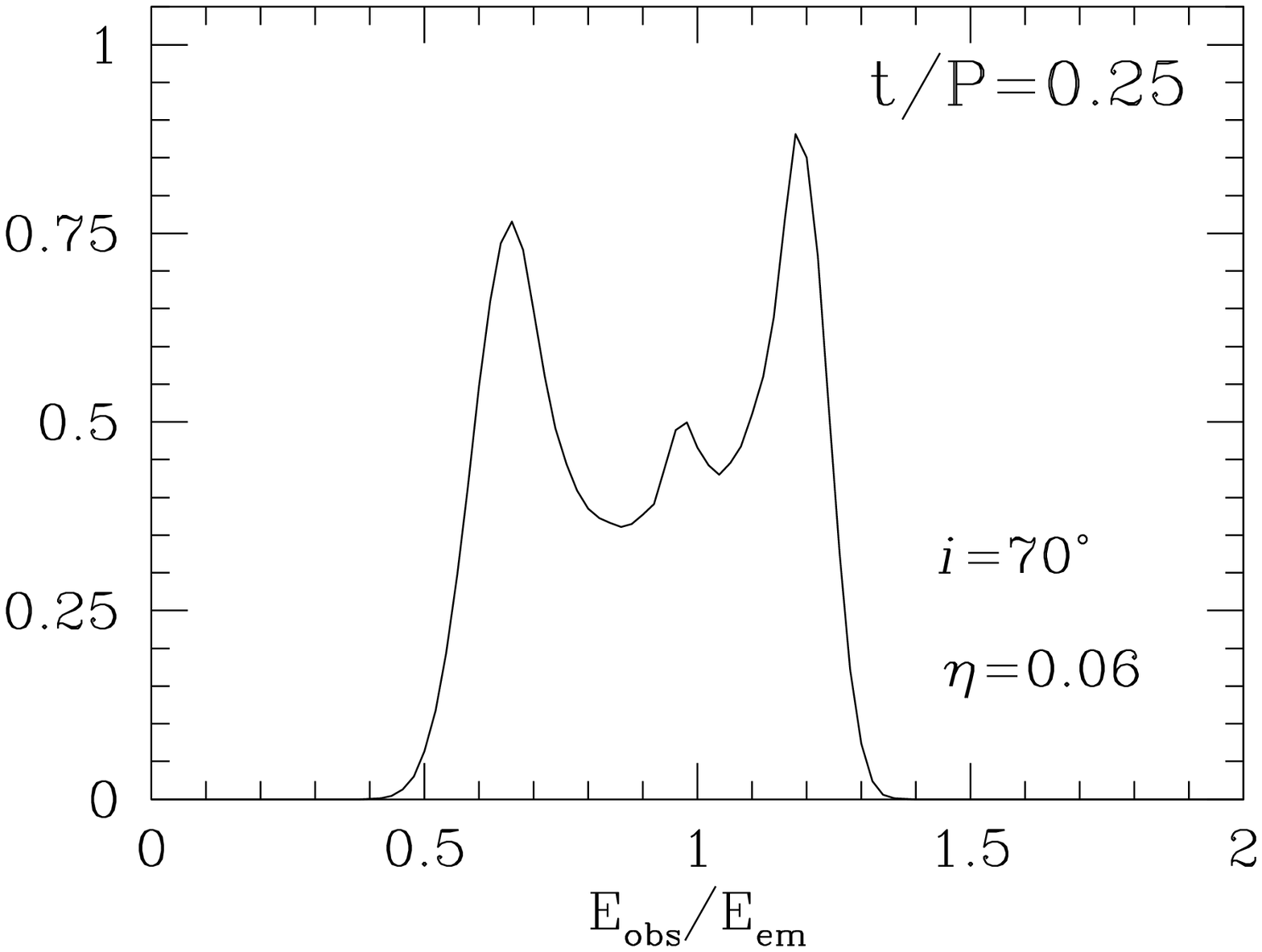}
\hskip 0.9truecm
\includegraphics*[width=5.cm]{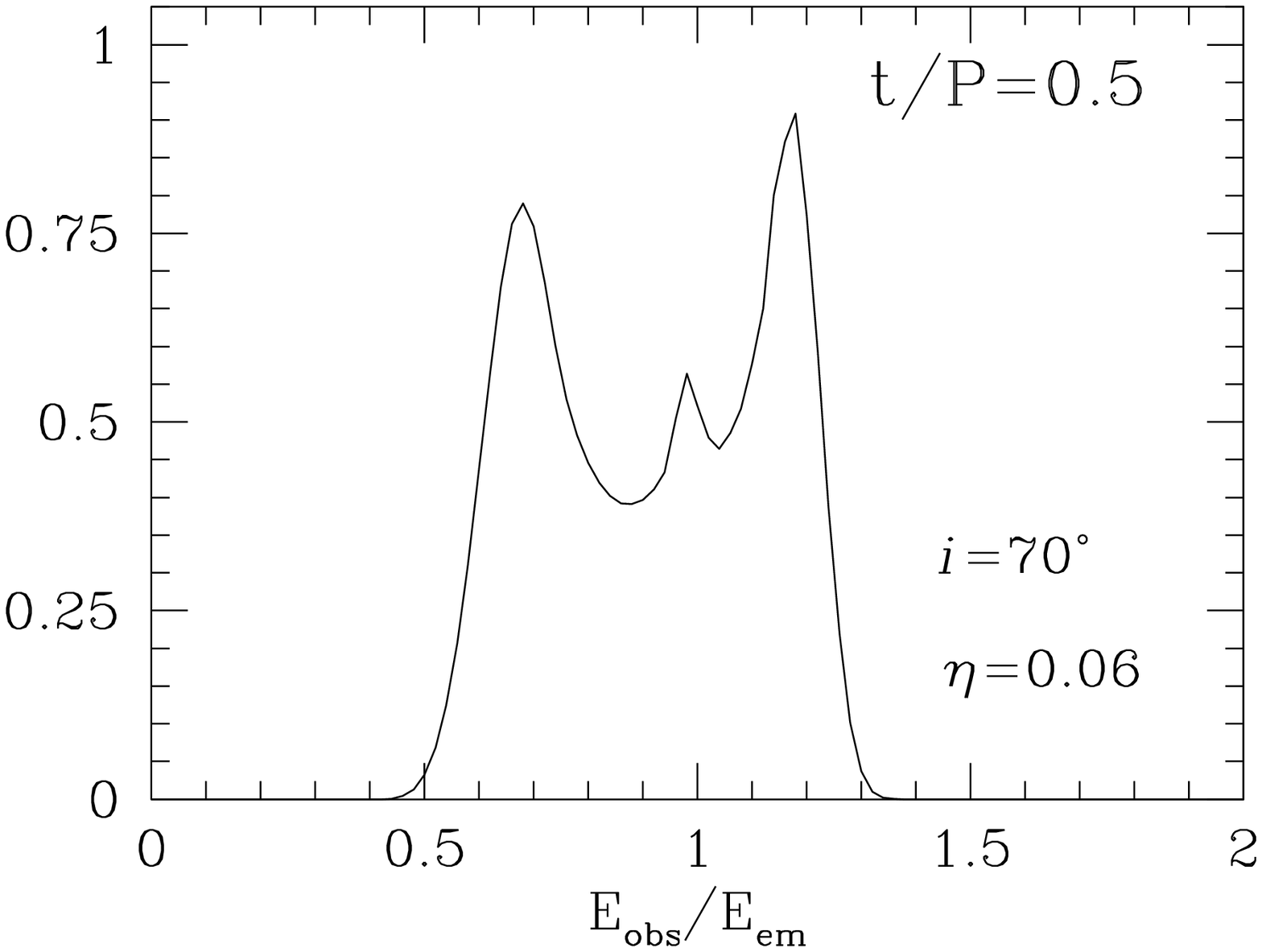}
\caption{\label{plottwo} \textit{Upper:} Ray-traced images of the
oscillating torus, shown at various phases of the fundamental $p$ mode,
with a logarithmic color scale of the X-ray intensity. \textit{Lower:}
Broadened emission line spectrum for each frame. The triple-peak spectrum
is caused by relativistic beaming towards and away from the observer,
along with a central peak due to gravitational lensing of the far side of
the accretion torus. The inclination is $i=70^\circ$ and the perturbation
amplitude is $\eta = 0.06$.}
\end{figure*}

	Figure~\ref{plotone} shows the light curves for a torus
orbiting around a Schwarzschild black hole of mass $M=10\;M_{\odot}$
at inclination angles of $10^\circ$ and $70^\circ$. The amplitude of
the initial perturbation was $\eta =0.02$, but qualitatively similar
behaviors are seen for perturbations in a linear regime, i.e.\
$\delta I/I \propto \eta$ for $\eta\lesssim 0.06$, where $I\equiv
\int^\infty_0 I_{\nu} d\nu$. Although 
it is clear that the light curve has a
quasi-periodic behavior and we find that this is strictly related to the
oscillating behavior of the torus at the same frequencies, the modulation
of the intensity is the combined result of several different relativistic
effects. In particular, for the optically thin emission model, the
minimum of the intensity is reached in the ``compression'' phase of the
oscillation, when the torus size is smaller and thus is closer to the
black hole. In this case, the gravitational red-shift reduces the
observed photon energies (as well as photon number through the invariance
of $I_\nu/\nu^3$) as the light has to escape from a deeper potential. At
the same time, the intensity is also varied by the special
relativistic beaming of photons emitted towards
and away from the observer. Finally, smaller contributions to the
intensity modulation also come from the transverse (``second-order'')
Doppler shift, and from the
gravitational lensing of the far side of the torus, magnifying a small
region of emission that is moving transverse to the
observer~\citep{beckw04,schni05b}.

	Because of the conservation of angular momentum during the
oscillations, the fluid velocities increase as the torus approaches the
black hole, thus enhancing the relativistic beaming of photons toward the
observer. For large inclination angles (i.e.\ when the torus is
almost ``edge-on''), this beaming is particularly intense and serves
to compensate for the intensity decrease due to the gravitational
red-shift. As a result, the two major relativistic effects counter each
other and the intensity modulation is smaller. For small inclination
angles (i.e.\ when the torus is almost ``face-on''), the beaming and
gravitational lensing can largely be ignored and the intensity
modulation is dominated by the gravitational red-shift and is thus
relatively larger (compare dashed and solid lines in
Fig.~\ref{plotone}). For these smaller inclination angles, simple
estimates can be made through the  
modulated red-shift $\delta\nu_{\rm obs} \equiv \nu_{\rm obs}
- \langle\nu_{\rm obs}\rangle$ and from the invariance of
$I_\nu/\nu^3$, to obtain $\delta I/I \approx 3(\delta\nu_{\rm
obs}/\nu_{\rm obs})$. The red-shift variation $\delta\nu_{\rm
obs}/\nu_{\rm obs}$ is in turn linearly proportional to $\eta$.

	It is important to underline that the dependence of the
modulation of the light curve on the inclination angle is the qualitative
opposite of the hot spot model described in~\citet{schni04} and could
serve to distinguish between the two models as more observations become
available. However, because this dependence is a function of the emission
model used, further studies are necessary.

	In the upper row of Figure \ref{plottwo} we show three snapshots
of an oscillating torus at different phases of a single
period $P$, with $t/P=0.0, 0.25, 0.5$, for a torus with inclination
angle $i=70^\circ$ and a perturbation amplitude $\eta =0.06$. In the
lower row, we show a broadened emission line spectrum for each frame with
the typical ``multi-horned'' features, where the two main peaks are
caused by photons emitted towards and away from the observer and thus
being blue- and red-shifted respectively. The smaller intermediate peak,
on the other hand, is due to the gravitational lensing of the far side of
the torus~\citep{beckw04,schni05b}.

	When the torus is more compact, closer to the black hole, and the
fluid velocities are comparatively larger, the observed emission line is
widest as the red-shifted wing moves to slightly lower energies and the
blue-shifted wing to slightly higher ones (see panel at $t/P=0.0$ in
Fig.~\ref{plottwo}). Furthermore, because of the relativistic invariance
of $I_\nu/\nu^3$, there are more blue-shifted photons, which increases
the total observed flux, compensating in part for the gravitational
red-shift of the smaller torus. On the other hand, when the torus is
at its maximum size and farther away from the black hole, the blue-
and red-shifts in the line spectrum are smaller, but the height of the
peaks is comparatively larger (see panel at $t/P=0.5$ in
Fig.~\ref{plottwo}). 

	To quantify more precisely the ``quasi-periodicity'' in the light
curve expected from an oscillating torus, we show in Figure
\ref{plotthree} the power spectra from the light curves reported in
Figure \ref{plotone}, in units of $[({\rm rms}/{\rm mean})^2 \mbox{
Hz}^{-1}]$.  The power is clearly dominated by a peak at $\sim 58$ Hz,
which coincides with the lowest $p$ mode of the torus (indicated as
the fundamental $f$). Furthermore, the power spectra also show smaller
peaks at the
overtones $o_1 \simeq 3/2 f$, $o_2 \simeq 2f$ and $o_3 \simeq 5/2f$, thus
demonstrating that the light curve possesses the same harmonic behavior
of the underlining hydrodynamics.

\vbox{
\vskip 0.125truecm
\centerline{\epsfxsize=7.5truecm 
\epsfbox{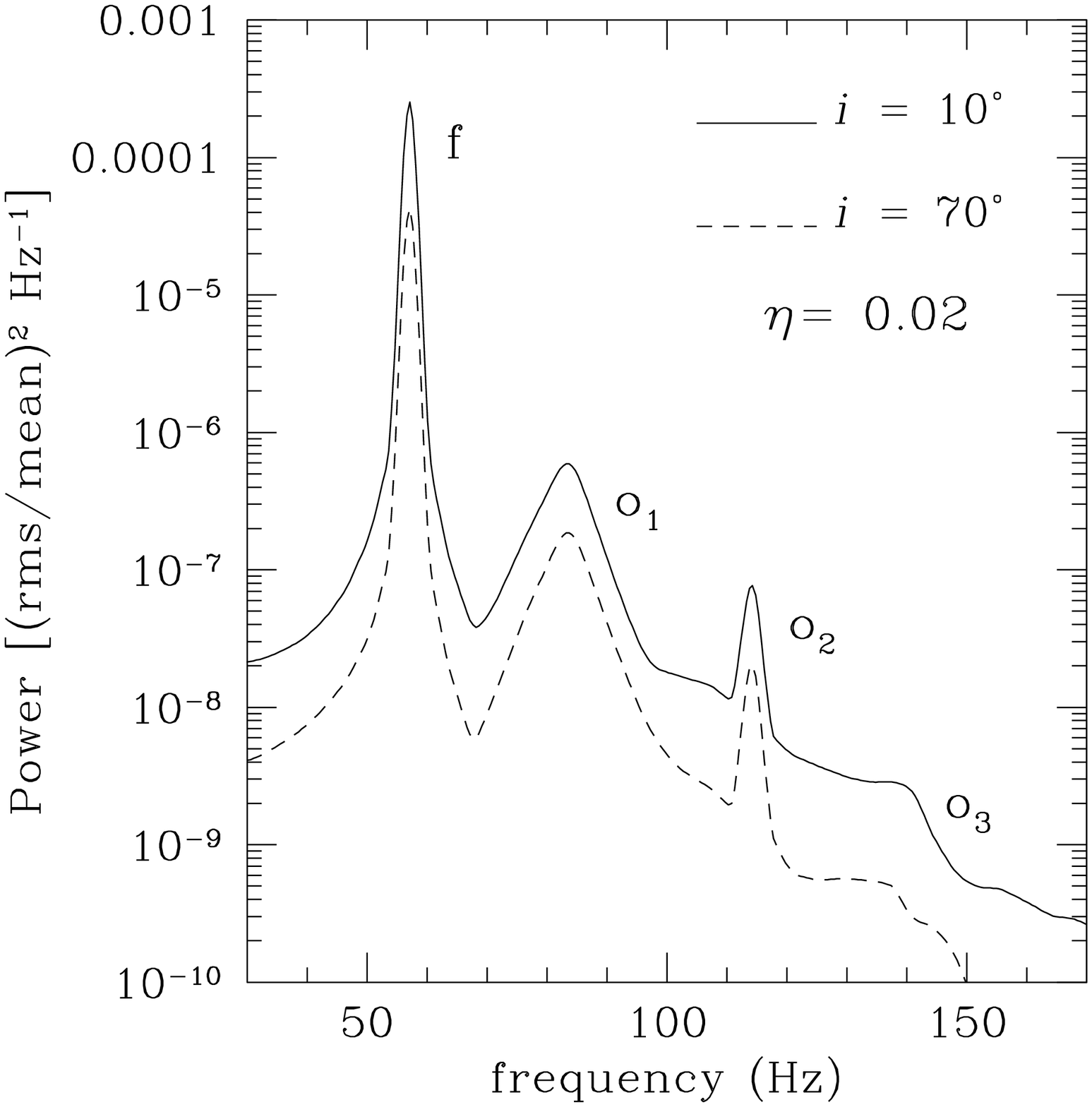}}

\figcaption[]{\label{plotthree}Power spectra of the light curves plotted
in Figure \ref{plotone}. Note the harmonic ratios among peaks, with $o_1
\simeq 3/2\, f$, $o_2 \simeq 2\, f$, and $o_3 \simeq 5/2\, f$.}
\vskip 0.125truecm
}

	To distinguish between transient modes and the more persistent
fundamental oscillations, we have also calculated power spectra for
longer time series, corresponding to more than 30 periods of
oscillations. In this case we find that the power in the higher modes is
reduced and may even disappear if the time series is restricted to the
later stages of the oscillations. This result, which is a numerical
artifact for low-amplitude perturbations, could however serve as a guide
in interpreting the observations. In a real accreting system, in fact, it
is likely that any perturbation lasts only a short time, as the turbulent
conditions of the accretion disk would continually create and destroy
coherent oscillating tori. As a result, these short-lived perturbations
could maintain more significant power in the higher harmonics than what
was produced with these simulations. Indeed, a characteristic torus
lifetime of only $3-5$ periods is sufficient to explain the observed
oscillator quality factors of $Q\approx 10-15$~\citep{schni05a}. Furthermore,
the discontinuous phase shift between subsequent tori provides then a
natural explanation for the broadening of the QPO peaks and the small
changes in frequencies observed as a ``jitter'' in the peaks.

\section{DISCUSSION AND CONCLUSIONS}
\label{discussion}

	We have demonstrated the positive correlation between the
intrinsic normal mode oscillations of a pressure-supported torus and the
extrinsic X-ray light curves and power spectra as seen by a distant
observer. This confirms the feasibility of the oscillating-torus model as
an explanation for the integer ratios seen in high-frequency QPO
peaks. The specific parameters of the torus model still require further
investigation in order to best fit the QPO data.

	For the simple emission model considered here, the variation in
the light curve is primarily caused by the gravitational red-shift of
photons coming from different radii as the torus moves in and out of the
black hole's potential well. Unlike the relativistic hot spot model, the
oscillating torus model predicts higher amplitude variations in the light
curve for smaller inclination angles, while at higher angles the special
relativistic beaming and gravitational lensing counter the gravitational
red-shift, reducing the variations in flux. This difference could be
crucial to distinguish the two models as more observations become
available.

	The results reported in this paper serve essentially to
demonstrate an issue of principle and have therefore been restricted to a
simplified scenario. Even with this advanced analysis tool, we
are still limited by the inherent uncertainty in the emission mechanisms
and the geometry of the surrounding accretion disk and corona. The fact
that the QPOs tend to appear exclusively in the Steep Power Law (``Very
High'') state of the black hole and are most significant in the higher
energy channels on {\it RXTE} \citep{mccli05} suggests that the specific
emission model may be constrained by the features of the X-ray energy
spectrum. These spectral features are also important in characterizing
the properties of the surrounding thermal disk and hot corona.

	Work is in progress to make the results presented here more
realistic by analyzing the dynamics of the torus in a Kerr space-time,
by considering electron scattering through a corona, and by including
optically thick line emission and thermal free-free emission with a
Kramer's opacity law. Initial calculations show qualitatively similar
results for the line emission model, but the thermal emission model
predicts a ``phase inversion'' with respect to the light curves shown in
this paper. This is due to the fact that for a barotropic equation of
state the temperature increases at smaller radii when the density and
pressure increase. This higher temperature produces greater X-ray
emission, outweighing the gravitational red-shift of the smaller
radius. These results will be presented in greater detail in a
forthcoming paper. 

\vspace{0.25cm}\noindent JDS is grateful to Edmund Bertschinger for his
insights and encouragement. Helpful discussions with Olindo Zanotti are
also acknowledged. Support comes by NASA grant NAG5-13306.


\end{document}